# Statistical Timing Analysis and Criticality Computation for Circuits with Post-Silicon Clock Tuning Elements

Bing Li and Ulf Schlichtmann, *Member, IEEE*

*Abstract*—Post-silicon clock tuning elements are widely used in high-performance designs to mitigate the effects of process variations and aging. Located on clock paths to flip-flops, these tuning elements can be configured through the scan chain so that clock skews to these flip-flops can be adjusted after manufacturing. Owing to the delay compensation across consecutive register stages enabled by the clock tuning elements, higher yield and enhanced robustness can be achieved. These benefits are, nonetheless, attained by increasing die area due to the inserted clock tuning elements. For balancing performance improvement and area cost, an efficient timing analysis algorithm is needed to evaluate the performance of such a circuit. So far this evaluation is only possible by Monte Carlo simulation which is very timing-consuming. In this paper, we propose an alternative method using graph transformation, which computes a parametric minimum clock period and is more than $10^4$ times faster than Monte Carlo simulation while maintaining a good accuracy. This method also identifies the gates that are critical to circuit performance, so that a fast analysis-optimization flow becomes possible.

## I. INTRODUCTION

PROCESS variations have become relatively larger in recent technology nodes. This trend makes the traditional worst-case timing analysis too pessimistic, leading to expensive overdesign and depriving designers of the valuable information of performance and yield. Modeling timing characteristics of a circuit more accurately, statistical static timing analysis (SSTA) has gained much attention in the research community in recent years [2]. This method represents process variations with random variables directly and computes the complete performance-yield curve, either in a parametric form or described by statistical properties, e.g., moments of different orders. Consequently, the yield of the circuit at any given clock period can be evaluated easily.

According to the assumption of the distributions of process variations, the method used to model gate delays, and the statistical operations in timing propagation, statistical timing algorithms can roughly be classified into several groups. First-order methods [3]–[5] use the canonical linear form [5] to represent gate delays and arrival times so that the recursive computations in timing analysis can be simplified but at the expense of accuracy. To improve modeling and propagation accuracy, quadratic methods are proposed in [6]–[10], using second-order polynomials to approximate gate delays and arrival times. Moreover, other methods, such as [11]–[13], can handle non-Gaussian delays and arrival times during timing propagation.

The research on statistical timing analysis focuses mainly on delay representation and arrival time propagation in combinational circuits and the resulting methods are implicitly applicable to circuits with flip-flops. In high-performance circuits, flip-flops with post-silicon clock tuning elements [14], [15] have also been deployed to counter process variations and improve circuit robustness. The tunable or programmable elements are inserted into the clock network to flip-flops that are relevant to critical paths. After manufacturing, the delay values of these elements are adjusted through the test access port to assign critical paths more timing budget by shifting the clock edges toward the stages with smaller delays. By allowing delay compensation across consecutive register stages, chips that might have failed to meet the timing specification can be revitalized. Therefore, with clock tuning elements the circuit can achieve a higher yield than without them.

Several methods have already been proposed for statistical timing analysis and optimization of circuits with clock tuning elements. In [16] a clock scheduling method is developed and clock tuning elements are selectively inserted to balance the skews due to process variations. Further in [17] algorithms are proposed to minimize the total area of these clock tuning elements, or to minimize the number of them in the circuit. In these methods, the yield of the circuit is computed using Monte Carlo simulation which consumes much runtime. In [18] the yield loss due to process variations and the total cost of clock tuning elements are formulated together for gate sizing. The resulting optimization problem is solved using a stochastic cutting-plane method with an STA scheme based on Monte Carlo simulation. This method still converges slowly due to the long runtime of yield evaluation. Moreover, the placement of clock tuning elements is investigated in [19] and a considerable benefit is observed when the clock tree is designed using the proposed tuning system.

The methods discussed above are applied as pre-silicon optimization or post-silicon adjustment before shipping the

A preliminary version of this paper was published as [1] in Proceeding of IEEE/ACM International Conference on Computer-Aided Design (ICCAD), 2011.

This work was supported in part by the German Research Foundation as part of the Transregional Collaborative Research Centre, Invasive Computing (SFB/TR 89).

Bing Li and Ulf Schlichtmann are with the Institute for Electronic Design Automation, Technische Universität München (TUM), Munich 80333, Germany (e-mail: b.li@tum.de; ulf.schlichtmann@tum.de).

chips to customers. Recently, further advances have been made to apply the clock tuning elements on-line to improve the lifetime performance [20]–[23]. The method in [21] adjusts the clock skews during runtime according to the occurrence of timing errors to achieve much better performance in timing-speculative circuits. The method in [23] explores the insertion of clock tuning elements and in-system configuration to reduce performance degradation due to aging. In addition, the work in [22] proposes an efficient post-silicon tuning method for each individual chip by searching a configuration tree combined with graph pruning. Moreover, the method in [20] applies clock tuning elements to compensate dynamic delay variations induced by temperature.

The research on statistical timing analysis and optimization has shown the advantage of using post-silicon clock tuning elements in high-performance designs. However, two issues still have not been addressed. The first is the need for a fast statistical timing analysis method, with which the runtime of the methods above can be reduced. Currently these methods use Monte Carlo simulation to compute the yield of the circuit considering process variations and are thus time-consuming.

The second issue is how to identify a set of critical gates for optimization of such a circuit containing post-silicon clock tuning elements. In statistical timing analysis the probability of a gate affecting the circuit performance is called criticality, and many methods have been proposed to describe and compute the criticalities of gates efficiently. In [5] the concept of criticality is first explored without considering correlation. In [24] the sensitivities of gate and path delays to the circuit delay are computed. In [25] the criticalities are computed using a cutset-based method combining with a binary tree partition. Furthermore in [26] a fast criticality computation method is proposed with incremental yield gradients. Additionally in [27] a clustering-based pruning is proposed to speed up the computation and improve the accuracy of criticalities. For ranking critical gates tiered criticalities are calculated to provide an order of statistical delays in [28], and for computing criticalities incrementally the reversible statistical max/min operation is investigated in [29]. These methods, though accurate and fast, do not consider post-silicon clock tuning elements, which allow the compensation of path delays across register stages but make the criticality computation more complicated.

In this paper, we propose a fast algorithm to evaluate the circuit performance in the presence of post-silicon clock tuning elements, so that yields of the circuit at different given clock periods can be calculated easily. Additionally, we investigate the criticalities of gate delays in the context that timing compensation is allowed across register boundaries. The registers considered in this paper are all edge-triggered flip-flops. The main contributions of this paper are as follows.

- The proposed method computes a parametric minimum clock period for the circuit with post-silicon clock tuning elements. The statistical properties of this minimum clock period, such as mean and variance, are directly available so that the yield of the circuit at any given clock period can be evaluated very fast. Since the computed circuit performance is in a parametric form, it can easily be integrated into other optimization methods that are built upon statistical timing analysis.
- The proposed method is much faster, more than $10^4$ times, than Monte Carlo simulation, by handling the path delay compensation across registers with a loop evaluation algorithm based on graph transformation.
- The criticalities of gate delays considering post-silicon clock tuning elements are defined and computed for circuit optimization. The proposed method can capture the critical gates within very short runtime, therefore enabling a fast analysis-sizing cycle.

The rest of this paper is organized as follows. In Section II we give an overview of the timing constraints considering delay tuning elements in circuits with edge-triggered flip-flops. These difference constraints are represented using a constraint graph in the formulation. In Section III the basic idea of calculating statistical minimum clock period from a constraint graph is defined and graph transformations are applied to extract it. Based on this result, the sequential criticality across flip-flop stages is defined to capture critical gates considering delay tuning elements. In Section IV, several implementation techniques are explained to accelerate the proposed algorithm. We discuss experimental results in Section V and conclude our work in Section VI.

## II. BACKGROUND AND PROBLEM FORMULATION

In this section, we describe the timing constraints of digital circuits with post-silicon clock tuning elements. These tuning elements can be configured after manufacturing to change the clock skews to flip-flops, according to path delays affected by process variations. Environmental variations are not considered in this method. In our formulation, all registers are edge-triggered flip-flops. Though transparent latches can also take advantage of these tuning elements, problem formulation becomes more complex in this case, because a timing constraint between a pair of latches contains more variables, so that the method discussed in this paper becomes inapplicable.

### A. Timing Constraints for Circuits with Post-Silicon Clock Tuning Elements

In high-performance digital designs, clock tuning elements are directly inserted into the clock paths to flip-flops. The intentional clock skews from these tuning elements are adjusted after manufacturing to counter process variations and aging effects [21]–[23]. These clock tuning elements may have various implementations and characteristics. The method in [30] produces tuning elements with precise adjustable delays shorter than 30 ps by way of voltage-controlled driver strength. The implementation in [15] uses a delay line and is capable of generating delays with 1 ps resolution. The de-skew buffers in [14] are built with CMOS inverters and arrays of passive loads, with 170 ps delay range and 8.5 ps step size. In [31] the delay element is made with a controlled contention circuit to provide a delay range around 140 ps with 8 steps. The delays of these tuning elements can be adjusted via test access port (TAP) after manufacturing [14].

Figure 1 illustrates an example of two flip-flops with clock tuning elements, where the clock signals $clk_i$ and $clk_j$ are not aligned anymore after passing the tuning elements, so that the timing budget allowed for the combinational circuit between flip-flops $i$ and $j$ can be regulated by the configurable delays $x_i$ and $x_j$. Assume that the clock signal switches at reference time 0. Then the clock events at flip-flops $i$ and $j$ happen at time $x_i$ and $x_j$, respectively. Therefore, a signal change at the output of flip-flop $i$ starts propagation at time $x_i$ and reaches flip-flop $j$ at the latest at time $x_i + \overline{d}_{ij}$, where $\overline{d}_{ij}$ is the maximum delay of the combinational circuit between $i$ and $j$. To guarantee the setup time constraint of $j$, this signal at the input of $j$ should be stable $s_j$ time before the rising clock edge of $j$, where $s_j$ is the setup time of $j$. Therefore, the timing constraint for the combinational circuit can be written as

$$x_i + \overline{d}_{ij} \leq x_j + T - s_j \tag{1}$$

where $T$ is the clock period. Let $\overline{w}_{ij} = \overline{d}_{ij} + s_j$. Then (1) is equivalent to

$$x_j - x_i \geq \overline{w}_{ij} - T. \tag{2}$$

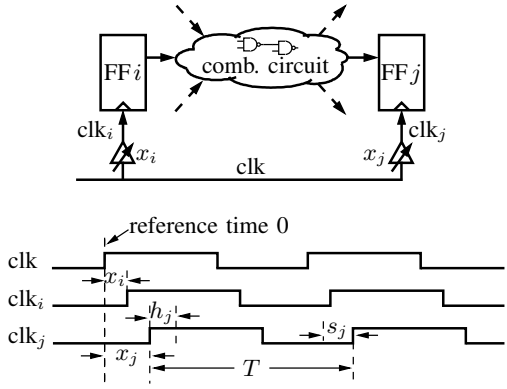

Fig. 1. Flip-flops with tuning elements. Clock signals $clk_i$ and $clk_j$ reach flip-flops unaligned due to the configurable delays $x_i$ and $x_j$. $s_j$ is the setup time of flip-flop $j$ and $h_j$ is the hold time of $j$. $T$ is the clock period.

Similar to setup time constraints, hold time constraints should also be included to guarantee that the signal propagated from the clock edge of flip-flop $i$ does not affect the latching function of flip-flop $j$ in the same cycle. Therefore, the tunable delays should satisfy the following constraint

$$x_i + \underline{d}_{ij} \geq x_j + h_j \tag{3}$$

where $\underline{d}_{ij}$ is the minimum combinational delay between $i$ and $j$; $h_j$ is the hold time of $j$. Let $\underline{w}_{ij} = h_j - \underline{d}_{ij}$. Then we can write (3) as

$$x_i - x_j \geq \underline{w}_{ij}. \tag{4}$$

In addition to the constraints (2) and (4), the maximum configurable delay or the tuning range is also limited due to area and power consumption [14], [15], [30], [31]. For a tuning element with delay $x_i$, its range is constrained as

$$0 \leq x_i \leq r_i \tag{5}$$

where $r_i$ is a constant representing the largest delay that the tuning element can add to the clock signal.

To guarantee the proper function of a circuit with clock tuning elements, the constraints (2), (4) and (5) are created for each pair of flip-flops between which there is a combinational path. Compared with the timing constraints of digital circuits without clock tuning elements, the constraints (2) and (4) contain additional variables $x_i$ and $x_j$ which establish the relation between the delays across register stages. In the timing constraints above, $\overline{d}_{ij}$ and $\underline{d}_{ij}$ can be calculated using the traditional breadth-first or depth-first propagation algorithms with statistical max and sum operations. However, in each constraint (2) or (4), the configurable delays $x_i$ and $x_j$ of the tuning elements can only be determined after manufacturing. In traditional static timing analysis, there are no such post-silicon tunable parameters. Therefore, by leaving out $x_i$ and $x_j$, the constraint (2) simply defines a lower bound for the clock period $T$, so that the minimum clock period of the circuit can be computed easily by calculating the maximum of all the lower bounds. But this method does not work any longer in the presence of the configurable delays $x_i$ and $x_j$. In addition, $\overline{w}_{ij}$ and $\underline{w}_{ij}$ are random variables, thus excluding the direct application of linear programming solvers to find the minimum clock period constrained by (2), (4) and (5), as in the classic clock skew optimization problem [32]. In addition, [33] provides a method to deal with discrete delay settings by adapting Bellman-Ford algorithm. This method works well with deterministic delays, but is still unable to solve the differential constraints when path delays become random variables due to process variations.

### B. Graph Representation of Difference Constraints

To establish a fast statistical timing algorithm for circuits with post-silicon clock tuning elements, we represent the constraints (2), (4) and (5) using a directed graph and calculate the statistical minimum clock period by graph transformation. In each of the constraints above, there are no more than two variables $x_i$ or $x_j$. Therefore, all these constraints together form a difference constraint problem [34], from which a constraint graph can be constructed. The constraint graph contains a node for each flip-flop, corresponding to a variable $x_i$ or $x_j$. If a constraint (2) exists for flip-flops $i$ and $j$, meaning that there is at least one combinational path from $i$ to $j$, a setup edge is created from node $i$ to node $j$ in the graph, with the weight $\overline{w}_{ij} - T$. Similarly, for the hold time constraint corresponding to (4) a hold edge is created from node $j$ to node $i$ with the weight $\underline{w}_{ij}$. To incorporate (5) into the constraint graph, a root node is created and shared by all the range constraints. The constraint itself can be split into $x_i \geq 0$ and $-x_i \geq -r_i$. For the former a range edge is created from the root node to node $i$ with the weight 0; for the latter a range edge from node $i$ to the root node with the weight $-r_i$.

In Fig. 2b the constraint graph of $s27$ from ISCAS89 benchmarks is illustrated as example. The nodes 1, 2 and 3 represent the three flip-flops in Fig. 2a. Edges between these nodes represent timing constraints (2) and (4), where only the weights of edges created from (2) contain $-T$. Node 0 is the shared root node. Edges to and from the root node correspond to the range constraints (5). In this example, we assume every flip-flop has a clock tuning element to show the basic idea. In reality, only those flip-flops that are relevant to critical paths are assigned tuning elements during circuit optimization [17]. In constructing the constraint graph, if a flip-flop has no tuning element, the corresponding variable $x_i$ or $x_j$ in (2) and (4) is fixed to 0 since no delay adjustment is possible. Consequently, the constraints (2) and (4) degrade into range constraints with only one variable, or additional yield constraints if both variables are set to 0.

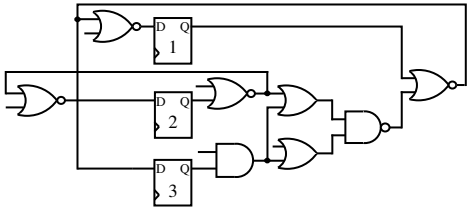

(a) s27 from ISCAS89 benchmarks without IO ports

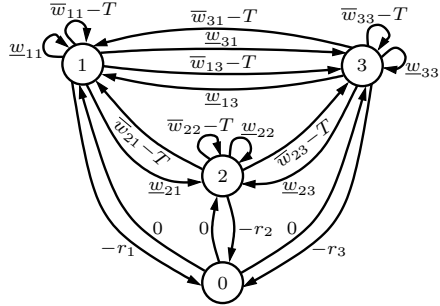

(b) Constraint graph of s27

Fig. 2. Construction of constraint graph. Nodes 1, 2 and 3 correspond to the flip-flops with clock tuning elements. Edges between these nodes represent timing constraints (2) and (4). Node 0 is the shared root node. Edges to and from the root node represent the range constraints (5).

The edge weights in the constraint graph contain random variables $\overline{w}_{ij}$ and $\underline{w}_{ij}$ defined in (2) and (4), respectively, so that linear programming can not be applied directly as in [32]. Another method to compute the minimum clock period is derived from the equivalence of difference constraints and non-positive loops in the constraint graph [34]. For a given clock period $T$, this equivalence specifies that a set of valid values for $x_i$ and $x_j$ that meets (2), (4) and (5) exists if and only if the sum of the edge weights from any loop in the constraint graph is no greater than 0. For example, the loop formed by the two setup edges and one hold edge across loop $2 \to 3 \to 1 \to 2$ in Fig. 2b should meet

$$(\overline{w}_{23} - T) + (\overline{w}_{31} - T) + \underline{w}_{21} \leq 0. \quad (6)$$

In the constraint graph, if all loops meet the conditions similar to (6), the given clock period $T$ can definitely be achieved by configuring the clock tuning elements. For a brief explanation, assume that all loops are non-positive. Under this condition, the Bellman-Ford algorithm [34] can always find the largest distances from the root node to all the other nodes in the graph. These maximum distances together form a valid configuration of all the tunable elements. For example, for a setup time constraint described by (2), we create an edge from node $i$ to node $j$ with the weight $\overline{w}_{ij} - T$ in constructing the constraint graph. The maximum distance from the root node to $j$ calculated by the Bellman-Ford algorithm is no smaller than the distance from the root node to $i$ plus the edge weight; otherwise the distance to $j$ is not the largest from the root node. That is to say, the maximum distances can naturally meet the constraint (2). This reasoning is valid for all the setup time constraints (2) and the hold time constraints (4). In addition, the range constraints (5) are guaranteed similarly by the range edges constructed from $x_i \geq 0$ and $-x_i \geq -r_i$ described earlier. Therefore, all the maximum distances together form a valid solution for the tunable delays.

Inversely, if there is a loop across which the sum of all edge weights is positive, the given clock period $T$ is infeasible. For example, if (6) is violated, we can deduce from (2) and (4) a contradiction as

$$0 = (x_3 - x_2) + (x_1 - x_3) + (x_2 - x_1) \geq \quad (7)$$
$$(\overline{w}_{23} - T) + (\overline{w}_{31} - T) + (\underline{w}_{21}) > 0. \quad (8)$$

In this case, the Bellman-Ford algorithm does not converge after sufficient iterations. Therefore a valid solution for the system of difference constraints (2), (4) and (5) requires that the loops in the constraint graph must be non-positive. Here we have only explained the basic idea about the equivalence between the non-positive loop condition and the existence of a solution for the difference constraint system. A detailed proof can be found in [34].

The discussion above shows that the non-positive loop condition can be used to verify whether a given clock period $T$ is feasible. This concept has been applied in [35] for clock schedule optimization, in [36] to determine the skew range in static timing analysis, in [37] for clock skew synthesis and in [38] for statistical timing verification of circuits using level-sensitive latches. In the following sections, we will explain a fast method based on parametric graph transformation and pruning techniques to calculate the statistical minimum clock period without enumerating all the loops in the graph.

## III. STATISTICAL TIMING ANALYSIS AND CRITICALITY COMPUTATION

In this section we first explain the concept of computing the statistical minimum clock period from the constraint graph of a circuit using parametric graph transformations in Section III-A and Section III-B. The basic idea of using these transformations was firstly introduced in [39] for statistical timing analysis of latch-controlled circuits. The challenges in applying these transformations to large graphs will be addressed by several techniques in Section IV. More importantly, we define *sequential criticality* considering timing compensation between sequential stages in Section III-C. This concept is a new layer of criticality above the criticality definitions in other works [5], [25], [27].

In the following discussion, we assume that each flip-flop has an individual clock tuning element, for simplification. The case that tuning elements are shared by multiple flip-flops can be modeled easily using the cluster method in [23].

### A. Defining $T_m$ Using Constraint Graph

In Section II we have discussed that the system of difference constraints formed by (2), (4) and (5) is equivalent to the condition that the constraint graph has no positive loops. In the case of static timing analysis, this condition can be verified using the Bellman-Ford algorithm for a given clock period. When process variations are considered, however, timing analysis becomes more complex because delays are represented by random variables. In the proposed method, we compute the statistical minimum clock period $T_m$ for such a circuit by graph transformation while keeping the clock period $T$ as an unknown variable. The resulting $T_m$ is a random variable from which the yield at any given clock period can be calculated easily.

For convenience, we write the edge weights in the constraint graph into a general form $w_{ij} - k_{ij}T$. For setup edges specified by (2), $k_{ij} = 1$ and $w_{ij} = \overline{w}_{ij}$; for hold edges specified by (4), $k_{ij} = 0$ and $w_{ij} = \underline{w}_{ji}$; for range edges (5), $k_{ij} = 0$ and $w_{ij} = 0$ or $w_{ij} = -r_i$. For hold edges we have switched the indexes so that the edge with a weight in the general form always has the direction from $i$ to $j$. Assuming that the clock period is still unknown, we can express the non-positive loop condition exemplified by (6) using edge weights in the general form as

$$w_l = \sum_{i,j}(w_{ij} - k_{ij}T) = \sum_{i,j} w_{ij} - \sum_{i,j} k_{ij}T \leq 0 \quad (9)$$

where $l$ is the index of the loop, $w_l$ is the weight of the loop, and the sum computation is applied over all edges on the loop.

According to the definition of $w_{ij} - k_{ij}T$, $k_{ij}$ is equal to 0 or 1. For loop $l$, if the sum of the coefficients $\sum_{i,j} k_{ij}$ is zero, there is no setup edge on the loop. In this case, (9) is a condition which only affects the yield of the circuit due to hold time constraints, but it has no effect on the minimum clock period. If $\sum_{i,j} k_{ij} > 0$, the loop $l$ contains setup edges and the constraint (9) specifies a lower bound for the clock period $T$ as

$$T_l = \sum_{i,j} w_{ij} \Big/ \sum_{i,j} k_{ij} \leq T \quad (10)$$

where $T_l$ is called loop constraint in the following discussion.

The constraint (10) from a loop creates a lower bound for the feasible clock period. If all loops in the constraint graph are considered, the minimum clock period $T_m$ for the circuit can be computed as

$$T_m = \max_{l \in L} T_l \quad (11)$$

where $L$ is the set of all loops in the constraint graph except those that only include hold edges or range edges. The loops of the latter type meet the condition $\sum_{i,j} k_{ij} = 0$ and are denoted by the set $\bar{L}$. Thereafter, the constraints (9) from these loops can be merged as

$$\max_{l \in \bar{L}} \{\sum_{i,j} w_{ij}\} \leq 0. \quad (12)$$

Because this variable only affects the yield of the circuit and its computation is similar to (11), we will focus only on the discussion of computing the minimum clock period $T_m$ in the following.

To compute the minimum clock period $T_m$ from all loops using (11) and the constraint in (12) directly requires that $T_l$ from every loop should be extracted. Obviously it is impractical to enumerate all these loops due to their prohibitive number in a large graph. In the following we will discuss three basic graph transformation operations to unroll loops gradually and extract the loop constraints in the form of $T_l$ at the same time.

### B. Computing $T_m$ Using Graph Transformation

Instead of enumerating all loops in the constraint graph, we compute $T_m$ in (11) using an iterative method based on graph transformation to capture the loop constraints $T_l$ in (10). Three basic graph transformation operations are used in the proposed method: self-loop removal, serial merge and parallel merge. Here a self-loop is formed by only one edge starting and ending at the same node, while a general loop may contain a chain of edges. Assume that the weight of the edge that forms a self-loop at node $i$ is $w_{ii} - k_{ii}T$. The non-positive loop constraint explained in the preceding section can be written as

$$w_{ii} - k_{ii}T \leq 0 \quad (13)$$

and thus be transformed to

$$w_{ii}/k_{ii} \leq T, \quad k_{ii} > 0 \quad (14)$$
$$w_{ii} \leq 0, \quad k_{ii} = 0. \quad (15)$$

For example, the self-loop with the weight $\overline{w}_{11} - T$ at node 1 in Fig. 2b requires that $T \geq \overline{w}_{11}$ while the self-loop with the weight $\underline{w}_{11}$ only constrains the yield. The self-loops are removed from the constraint graph right away once they appear and the corresponding constraints are merged in (11) and (12), respectively.

The self-loops are removed and need not to be considered again in capturing the constraints from other loops, because the extracted lower bounds in the form of (14) or (15) guarantee that the weight of the edge in a self-loop is not positive, so that the weights of other loops including this edge can not increase compared with the case that the self-loop is not included. For example, in Fig. 2b there is a loop formed by the setup edge from node 3 to node 1, then the self-loop with weight $\overline{w}_{11} - T$, then the hold edge from 1 to 2 and the setup edge from 2 to 3. If the clock period $T$ is no less than $\overline{w}_{11}$, the timing constraint is guaranteed implicitly if the constraint from the loop without the edge forming the self-loop can be met.

After removing self-loops, a typical structure in the constraint graph is illustrated on the left side of Fig. 3. The serial merge operation removes node $v$ from this structure and creates direct edges between each predecessor and each successor of node $v$. The weight of a new edge is equal to the sum of the weights of the edges from which the new edge is constructed. Therefore, the constraint from any loop that passes through node $v$ is not affected. The new weight is also in the general form $w_{ij} - k_{ij}T$, so that the serial merge operation can be applied iteratively.

In the serial merge operation, a predecessor node and a successor node may be the same. For example, if we apply the serial merge operation to node 1 in Fig. 2b after all original self-loops are removed, a new self-loop is constructed at node 2 due to the setup edge from node 2 to node 1 and the hold edge from 1 to 2, with edge weight $\overline{w}_{21} - T + \underline{w}_{21}$. This self-loop is immediately removed from the graph during the graph transformation and the timing constraint $T \geq \overline{w}_{21} + \underline{w}_{21}$ is captured and merged to $T_m$ using (11). Because the serial merge operation creates direct edges between predecessor and successor nodes iteratively, the newly created and then removed self-loops actually contain the sum of the edge weights from the original constraint graph. In other words, the original loops in the graph are collapsed by graph transformation and their loop constraints are captured by self-loops eventually.

After each serial merge operation the number of nodes in the constraint graph is reduced by 1, but many new edges may be created in the graph, because in the worst case $m \times n$ new edges could be created for the node $v$ with $m$ predecessors and $n$ successors. Usually this is far larger than the number of the removed $m + n$ edges, thus causing the edge number in the graph to increase very quickly during the transformation. Actually applying the serial merge operation repeatedly to cap-

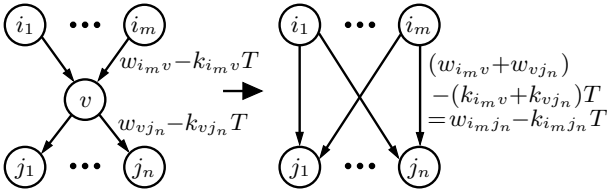

Fig. 3. Serial merge operation. Direct edges are created after node $v$ is removed. New edge weights are calculated as the sums of the former weights.

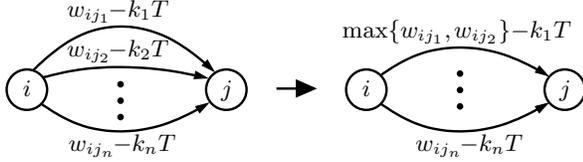

Fig. 4. Parallel merge operation. Edges connecting the same nodes and having the same coefficients of $T$ are merged. The new edge weight is computed as the maximum of the former edge weights. In this case $-k_1$ is equal to $-k_2$.

**Algorithm 1:** Computing the minimum clock period $T_m$ from the constraint graph using graph transformation

- L1 $G$: the constraint graph created from (2), (4) and (5);
- L2 $\nu$, $\nu_i$, $\nu_j$: nodes involved in graph transformation operations;
- L3 $N$: the number of nodes in the original constraint graph.
- L4 **foreach** *node $\nu$ in the constraint graph $G$* **do**
- L5     remove_self_loops($\nu$);
- L6     update_$T_m$ ();
- L7 **end**
- L8 **for** $k$=1 **to** $N$ **do**
- L9     prune_edges($G$);
- L10     $\nu$=select_node($G$);
- L11     serial_merge($\nu$);
- L12     **foreach** *predecessor node $\nu_i$ of $\nu$* **do**
- L13         remove_self_loops($\nu_i$);
- L14         update_$T_m$ ();
- L15         **foreach** *successor node $\nu_j$ of $\nu$* **do**
- L16             **if** *there exist parallel edges connecting nodes $\nu_i$ and $\nu_j$* **then**
- L17                 parallel_merge($\nu_i$, $\nu_j$);
- L18             **end**
- L19         **end**
- L20     **end**
- L21 **end**

ture all loop constraints without further pruning is equivalent to enumerating all the loops in the graph directly, which is very time-consuming for a large constraint graph. To solve this problem, we will apply various pruning techniques to be discussed in Section IV to reduce the number of edges in each iteration.

Besides the serial merge operation, we apply the parallel merge operation to reduce the number of edges further. In the constraint graph, if there are multiple edges between two nodes, these edges are called parallel edges. For example, between nodes 1 and 3 in Fig. 2b there are two sets of parallel edges. Additionally, parallel edges may also be created by the serial merge operation. For example, if node 1 in Fig. 2b is removed by the serial merge operation, new parallel edges appear between nodes 2 and 3. These new edges may also have the same coefficient $k_{ij}$ of $T$, so that they can be merged to reduce the number of edges, by the operation called parallel merge and illustrated in Fig. 4. If the coefficients $-k_1$ and $-k_2$ of $T$ in the weights of the two parallel edges in Fig. 4 are equal, the first two edges can be merged into one edge, whose weight is computed as the maximum of two former edge weights. If these two coefficients are not equal, we can not merge the two parallel edges because $T$ is kept as an unknown variable to catch its lower bounds constrained by loops so that we can not compare the weights of the two edges directly. Similar to the serial merge operation, parallel merge operation does not affect the constraints from the weights of loops that pass through $i$ and $j$, because the maximum weight of the loops through the merged edges is maintained by the new edge.

Applying the graph operations discussed above iteratively we can capture all loop constraints and compute the minimum clock period $T_m$ by Algorithm 1. At the beginning, the algorithm removes all original self-loops in L4–L7 to reduce the number of edges. In each of the following iterations, a node is removed using the serial merge operation denoted by serial_merge($\nu$) where $\nu$ is the node selected by the function select_node($G$) from the constraint graph $G$. The function remove_self_loops($\nu$) removes edges that form self-loops at node $\nu$. These self-loops may exist in the original constraint graph or are results from merging the edges on the loops by iterative serial merge operations. For each of these self-loops the constraint in the form of (10) is computed and $T_m$ is updated by the function update_$T_m$ () using (11). After each serial operation, only checking the nodes that are the predecessors of the removed node at L12 is enough, since a self-loop can only be formed when the predecessor and successor of the removed node are the same. After each serial merge operation, parallel merge operations are applied to compress edges in the graph further. The algorithm needs to run $N$ iterations, where $N$ is the number of the nodes in the original constraint graph. The constraints from all the loops are captured when eventually all of the nodes are deleted.

Algorithm 1 only shows the basic concept of applying graph transformation operations. In spite of the self-loop removal and parallel merge operations, the number of edges in the constraint graph may still increase very fast. To solve this problem, we apply pruning techniques denoted by the function prune_edges($G$) and discuss them as well as the function select_node($G$) used to select the next candidate for the serial merge operation in Section IV.

### C. Computing Criticalities Considering Clock Tuning Elements

For circuit optimization the timing analysis tool should report a set of gates that are critical to the circuit performance. The probability that a gate delay affects the circuit performance is called criticality [5], [25], [27]. Because clock tuning elements allow the path delays to compensate each other across flip-flops, critical paths in such circuits may span more than one stage. An example of these critical paths is shown in Fig. 5, where the inverters represent combinational paths with delays above them. The ranges of the clock tuning elements are 3. In this example, we use deterministic delays to show the basic idea. In real circuits, process variations expand the delays to statistical distributions, which we discuss right after explaining this example.

In Fig. 5, if the clock tuning elements are not considered, the critical path is between flip-flops 1 and 2. However, the clock tuning element at flip-flop 2 allows a minimum clock period of 5 at this stage if the intentional skew is configured to 3. On the contrary, the minimum clock period constrained by the paths between flip-flops 4 and 6 is still 6, because the change of the clock skew to flip-flop 5 invariably increases the clock period constrained by one of these two combinational paths. Therefore, the paths between flip-flops 4 and 6, though not having the largest combinational delay in the circuit, become the critical paths.

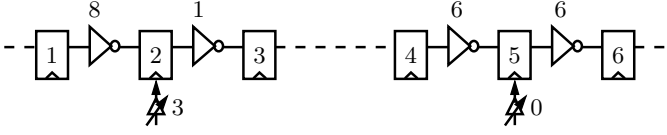

Fig. 5. Critical paths in the presence of clock tuning elements whose tuning ranges are up to 3. The paths with the delays 6 instead of the path with the delay 8 are critical.

To capture the critical paths in the presence of clock tuning elements, we first define the criticalities for loops and edges in the constraint graph. Thereafter, the concept in [25], [26] is extended to include this information into the computation of criticalities for combinational gates. Note here other criticality computation methods such as [27] can also be extended by incorporating the loop constraints similarly.

According to (11) the circuit performance is constrained by the loop constraints $T_l$ from all the loops in the constraint graph. Because edge weights are random variables in real circuits, any loop has a probability to dominate the circuit performance. The probability that a loop $l$ is critical is thus defined as

$$c_l = prob\{T_l \geq T_m\} \quad (16)$$

where $T_m$ is defined in (11). Because $T_m$ is the maximum of the loop constraints, the definition of $c_l$ in (16) is the probability that the loop constraint $T_l$ is no smaller than any constraints from other loops. The larger $c_l$ is the more the loop $l$ affects the circuit performance. Therefore $c_l$ is called loop criticality for loop $l$ and a loop with a large $c_l$ is a critical loop. The edges on a critical loop are candidates for optimization.

In the constraint graph an edge may be on multiple loops. If any of these loops dominates the circuit performance, the edge is critical. Therefore for an edge $e$ representing the combinational delay between a pair of flip-flops, the *sequential criticality* is defined as

$$c_e = prob\{\bigvee_{l \in L_e} (T_l \geq T_m)\} \quad (17)$$

$$= prob\{\neg(\bigwedge_{l \in L_e} (T_l < T_m))\} \quad (18)$$

$$= prob\{\neg(\max_{l \in L_e}\{T_l\} < T_m)\} \quad (19)$$

$$= prob\{\max_{l \in L_e}\{T_l\} \geq T_m\} \quad (20)$$

where $L_e$ is the set of loops across $e$ and $\max_{l \in L_e}\{T_l\}$ is the maximum of the constraints from all these loops. $\wedge$ means *logic and*, $\vee$ *logic or* and $\neg$ *logic not*.

An edge in the constraint graph, if not connected to the root node, corresponds to a combinational delay between a pair of flip-flops in the circuit. In the following, we will discuss the definition of criticalities for combinational gate delays considering the effect of clock tuning elements. The discussion will focus on maximum combinational delays constrained by setup time constraints (2). The criticalities corresponding to hold time constraints (4) can be computed similarly.

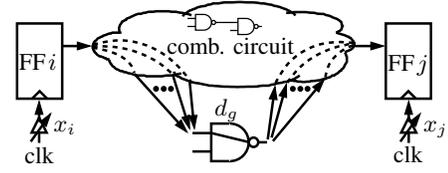

Fig. 6. Path partition for computing criticalities of combinational delays [25], [26]. The gate delay is critical if the paths across it dominate the other paths.

The sequential criticality above indicates whether the maximum delay of combinational paths between a pair of flip-flops is critical. In statistical timing analysis, it is often required to calculate the probability that a gate is on a critical path for circuit optimization, as in [5], [25], [27]. In the following, we explain how the sequential criticality is incorporated into the definition of criticalities for gate delays, using the cut-set concept in [25], [26] as an example. Note that the proposed sequential criticality is an additional layer above the existing definitions of criticality, and it can be combined with any of the methods in [5], [25], [27].

The cutset concept in [25], [26] is introduced to define the criticality of a gate in a combinational circuit. This concept is illustrated in Fig. 6, where the paths in the combinational circuit between flip-flops $i$ and $j$ can be partitioned into two sets: $P_g$ and $P_{\bar{g}}$. $P_g$ contains the paths going through the gate $g$; $P_{\bar{g}}$ contains all the other paths between $i$ and $j$. The gate delay is critical if the path with the maximum delay passes it, that is to say, the critical path belongs to $P_g$. The maximum path delay $d_{P_g}$ from $P_g$ can be computed by

$$d_{P_g} = d_{ig} + d_g + d_{gj} \quad (21)$$

where $d_{ig}$ is the maximum delay from $i$ to the input of $g$, and $d_{gj}$ is the maximum delay from the output of $g$ to $j$. $d_g$ is the gate delay. Therefore, the probability that the gate delay is critical can be defined as

$$c_g^{no\_tuning} = prob\{d_{P_g} \geq d_{P_{\bar{g}}}\} \quad (22)$$

$$= prob\{d_{P_g} \geq \max\{d_{P_g}, d_{P_{\bar{g}}}\}\} \quad (23)$$

$$= prob\{d_{P_g} \geq d_{ij}\} \quad (24)$$

where $d_{P_{\bar{g}}}$ is the maximum path delay from $P_{\bar{g}}$; $d_{ij}$ is the maximum delay of all paths between $i$ and $j$. Note the equations (22)–(24) are valid only under assumption of an exact statistical maximum function [26].

In the circuit, the gate $g$ may be on the combinational paths between many pairs of flip-flops, corresponding to a set of edges, written as $E_g$, in the constraint graph. For example, in Fig. 2a, the NOR gate connected directly to the output of flip-flop 2 is on the combinational paths from flip-flop 2 to flip-flops 1, 3 and itself, respectively. If any of these paths is critical, the gate is a critical gate. By combining (17)–(20) and (24), the criticality of a gate delay in the presence of clock tuning elements is defined as

$$c_g = prob\{\bigvee_{e \in E_g} (\max_{l \in L_e}\{T_l\} \geq T_m \wedge d_{P_g} \geq d_{ij})\} \quad (25)$$

$$= prob\{\bigvee_{e \in E_g} (0 \geq \max\{T_m - \max_{l \in L_e}\{T_l\}, d_{ij} - d_{P_g}\})\} \quad (26)$$

**Algorithm 2:** Computing criticalities of gate delays by edge tracing

- L1 $G$: the constraint graph created from (2), (4) and (5);
- L2 $\nu, \nu_i, \nu_j$: nodes involved in graph transformation operations;
- L3 $\epsilon, \epsilon_i, \epsilon_j$: edges involved in criticality computation;
- L4 $N$: the number of nodes in the original constraint graph.
- L5 Calculate $T_m$ using Algorithm 1;
- L6 **foreach** *node $\nu$ in the constraint graph $G$* **do**
- L7     remove_self_loops($\nu$);
- L8     **foreach** *removed edge $\epsilon$* **do**
- L9         update_loop_constraint($\epsilon$);
- L10     **end**
- L11 **end**
- L12 **for** $k=1$ *to* $N$ **do**
- L13     prune_edges($G, T_m$);
- L14     $\nu$=select_node($G$);
- L15     serial_merge($\nu$);
- L16     **foreach** *new edge $\epsilon$ created from the successive edges $\epsilon_i$ and $\epsilon_j$ by serial merge operation* **do**
- L17         update_tracing_lists($\epsilon, \epsilon_i, \epsilon_j$);
- L18     **end**
- L19     **foreach** *predecessor $\nu_i$ of $\nu$* **do**
- L20         remove_self_loops($\nu_i$);
- L21         **foreach** *removed edge $\epsilon$ in a self-loop* **do**
- L22             **foreach** *edge $\epsilon_i$ in the tracing list of $\epsilon$* **do**
- L23                 update_loop_constraint($\epsilon_i$);
- L24             **end**
- L25         **end**
- L26         **foreach** *successor node $\nu_j$ of $\nu$* **do**
- L27             **if** *there exist parallel edges connecting nodes $\nu_i$ and $\nu_j$* **then**
- L28                 parallel_merge($\nu_i, \nu_j$);
- L29             **end**
- L30         **end**
- L31     **end**
- L32 **end**
- L33 **foreach** *edge $\epsilon$ with sequential criticality $c_\epsilon > 0$* **do**
- L34     compute_combinational_gate_criticality ();
- L35 **end**

where edge $e$ is between nodes $i$ and $j$ in the constraint graph, and $L_e$ is the set of loops containing edge $e$ in the graph.

Let $\mathcal{C}_e = \max\{T_m - \max_{l \in L_e}\{T_l\}, d_{ij} - d_{P_g}\}$. Then (26) can be rewritten as

$$c_g = prob\{\bigvee_{e \in E_g}(0 \geq \mathcal{C}_e)\} \quad (27)$$

$$= prob\{\neg(\bigwedge_{e \in E_g}(\mathcal{C}_e > 0))\} \quad (28)$$

$$= prob\{\neg(\min_{e \in E_g}\{\mathcal{C}_e\} > 0)\} \quad (29)$$

$$= prob\{\min_{e \in E_g}\{\mathcal{C}_e\} \leq 0\}. \quad (30)$$

To compute the criticalities, several variables in (17)–(20) and (25)–(30) should be known. The minimum clock period $T_m$ can be computed using Algorithm 1. The path delays $d_{ij}$ and $d_{P_g}$ can be computed using an SSTA engine as in [25], [26]. The computation of $\max_{l \in L_e}\{T_l\}$ for an edge $e$ in the constraint graph needs to trace the loops containing the edge $e$ and is explained in the following by extending Algorithm 1.

In Algorithm 1 the function remove_self_loops() deletes self-loops and updates $T_m$ with the newly computed $T_l$ using (11). These loops are either in the original graph, or created from multiple edges by the serial and parallel merge operations. In Algorithm 1, we only keep the weights of the edges during the iterations. Consequently, we have no information to identify the original edges from which a new lower bound $T_l$ is generated, so that we can not update $\max_{l \in L_e}\{T_l\}$ for individual edges. To solve this problem, we maintain an edge tracing list for each new edge to trace the edges from which the new edge is created. When two consecutive edges are replaced by a new edge during the serial merge operation, the edge tracing lists maintained for the two replaced edges are combined together to construct the new edge list. Because the edges in the graph may have different edge tracing lists, the parallel merge operation in Fig. 4 can not be applied directly even if the edge weights have the same coefficients of $T$, because we need to keep separate edge records so that we can trace back to the original edges when new self-loop constraints are extracted. This limitation increases the number of edges during the graph transformation and leads to a long runtime. In Section IV, we will explain how to adapt the parallel merge operation to handle edge tracing lists. In the iterations each time when a self-loop is formed, the loop is removed and the loop constraint $T_l$ defined in (10) is updated into the random variable holding $\max_{l \in L_e}\{T_l\}$ for each edge in the edge tracing list. After the iterations are finished, all loop constraints are extracted and the loop constraint $\max_{l \in L_e}\{T_l\}$ for each original edge is computed, so that the criticality for gate delays defined in (30) can be calculated.

The basic concept of criticality computation is summarized in Algorithm 2. The main structure of this algorithm is similar to that of Algorithm 1 because they both capture the non-positive constraints from loops. The difference is that Algorithm 1 updates the extracted constraints into $T_m$ by the function update_$T_m$() using (11), but in L8–L10 and in L21–L25 Algorithm 2 updates the constraints into the random variables holding $\max_{l \in L_e}\{T_l\}$ for individual edges that contribute to the weight of the removed self-loop, using the function update_loop_constraint(). After the iterations L12–L32 are finished, we have the loop constraints $\max_{l \in L_e}\{T_l\}$ for all the edges. Thereafter, the criticalities for gate delays are calculated using (30) in L33–L35, where $d_{ij}$ and $d_{P_g}$ are calculated for the combinational paths corresponding to edge $e$. In this process, we only need to consider the edges with sequential criticality $c_e$ larger than 0, because other edges are dominated by these edges and thus do not affect the minimum clock period.

## IV. ACCELERATION TECHNIQUES AND DISCUSSIONS

In computing the minimum clock period and criticalities, the edges in the constraint graph are transformed as shown in the main iterations of Algorithm 1 and 2. By connecting the predecessors and successors of a node directly, the serial merge operations in the iterations actually unroll all loops to capture the non-positive constraints. Without further improvements, these algorithms may be very time-consuming due to the large number of loops in the constraint graph. In this section, we discuss implementation techniques to enhance Algorithm 1 and 2 for better performance.

*1) Eliminating edges using ranges of clock tuning elements*

The first technique we use is to remove the edges that are dominated by other edges in the constraint graph. These edges have small weights so that any configuration of the clock tuning elements does not make them critical. This idea has been used in [22] for post-silicon configuration. In this paper, we extend it to handle statistical delays and apply it during iterations to process edges formed by the serial merge operations. These new edges represent the concatenated edges on the paths in the original constraint graph, so that they are potential candidates to be pruned due to the delay compensation across multiple flip-flop stages.

Consider the edge between flip-flops 2 and 3 in Fig. 5, where the ranges of the tuning elements are 3. In any delay configuration of these tuning elements, the path connecting these flip-flops does not affect the minimum clock period, so that we can remove it before starting the graph transformation. Now we consider another example in Fig. 2b. If the weight $\overline{w}_{23} - T$ of the setup edge from node 2 to node 3 is smaller than $-r_2$, the setup time constraint from this edge can never be violated, because the largest possible configurable clock skew to flip-flop 2 is $-r_2$ and the smallest one to flip-flop 3 is 0. In the constraint graph, the edge from node 2 to the root node with the weight $-r_2$ is created according to the former range constraint, and the edge from the root node to node 3 with the weight 0 is created according to the latter range constraint. If $\overline{w}_{23} - T < -r_2$ holds as discussed above, the setup edge is always dominated by the concatenated edges from node 2 to the root node with the weight $-r_2$ and from the root node to node 3 with the weight 0. Therefore, the setup edge actually makes no contribution to the minimum clock period. Consequently, we can remove this edge safely without losing any accuracy in statistical timing analysis. In the general case, an edge with the weight $w_{ij} - k_{ij}T$ can be removed if it meets
$$w_{ij} - k_{ij}T < -r_i \quad (31)$$
where $r_i$ is the largest configurable delay of the clock tuning element attached to the source node $i$ of the edge.

In the condition (31), $w_{ij}$, $k_{ij}$ and $r_i$ are already known. Because a feasible clock period $T$ should always be no less than the minimum clock period $T_m$, that is, $T \geq T_m$, we check the condition (31) using
$$w_{ij} - k_{ij}T_m < -r_i \quad (32)$$
which is a sufficient condition of (31). In the iterations of Algorithm 1, $T_m$ increases gradually when new loop constraints are merged into it using (11). Therefore, the edge elimination technique is more effective in pruning edges in the later iterations of Algorithm 1 and in the computation of criticalities in Algorithm 2. Since both $w_{ij}$ and $T_m$ are random variables, the condition (32) can only hold with a probability as
$$prob\{w_{ij} - k_{ij}T_m < -r_i\}. \quad (33)$$
If this pruning probability for an edge approximates 1, for example, if it is larger than 0.98 as in our experiments, we remove the edge from the constraint graph to reduce runtime. This technique is implemented in the function prune_edges() in Algorithm 1 and 2.

*2) Pruning edges by parallel dominance*

In the serial merge operation in Fig. 3, direct edges are created between the predecessors and successors of the removed node. After many iterations, a large number of parallel edges may appear. But the parallel merge operation in Fig. 4 can only handle parallel edges with the same coefficients of $T$. To reduce the number of edges in the constraint graph further, we compare the weights of parallel edges and remove the edge whose weight is dominated by the other edges.

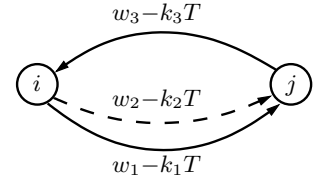

Fig. 7. Parallel pruning. The edge with the weight $w_2 - k_2T$ can be removed if it is statistically dominated by the edge with the weight $w_1 - k_1T$. This removal shall not affect the accuracy of $T_m$ and the lower bounds $T_l$ in (17)–(30) of the edges that are on the loops passing the dominated edge to guarantee correct criticalities.

Consider two edges with the weights $w_1 - k_1T$ and $w_2 - k_2T$ where $k_1 \neq k_2$. If the weight of the second edge is dominated by the weight of the first edge, the second edge is removed from the constraint graph. The removal condition can be expressed as
$$w_1 - k_1T > w_2 - k_2T. \quad (34)$$
In case of $k_2 - k_1 > 0$, we can transform (34) into
$$T > (w_2 - w_1)/(k_2 - k_1). \quad (35)$$
Similar to the computation in (31)–(33), we substitute $T_m$ for $T$ to create a sufficient condition of (35) as
$$T_m > (w_2 - w_1)/(k_2 - k_1). \quad (36)$$
Because edge weights are random variables, the comparison can only be performed statistically. Therefore we compute the probability of parallel dominance as
$$prob\{T_m > (w_2 - w_1)/(k_2 - k_1)\}. \quad (37)$$
If this probability is close to 1, the second edge is dominated by the first edge so that it can be removed from the constraint graph.

The parallel pruning technique removes edges whose delays are statistically dominated by others. In the following, we explain this technique with more details. In the example illustrated in Fig. 7, two parallel edges form loops with the edge having the weight $w_3 - k_3T$ representing the paths from node $j$ to node $i$. Assume that the edge with the weight $w_2 - k_2T$ is dominated by the edge with the weight $w_1 - k_1T$ and removed from the graph. Furthermore, assume that in the current iteration in Algorithm 1 the lower bound of $T$ created by the extracted loop constraints is $T_c$, which is the current value of $T_m$ before the loop constraints from the two loops in Fig. 7 are processed. Let $T_1 = (w_1 + w_3)/(k_1 + k_3)$ and $T_2 = (w_2 + w_3)/(k_2 + k_3)$. If only the constraint from the dominating edge is extracted, the new constraint for $T$ can be written as
$$T \geq \max\{T_c, T_1\} = T'. \quad (38)$$
Thereafter, if the constraint from the dominated edge would be included, the constraint for $T$ becomes
$$T \geq \max\{T', T_2\}. \quad (39)$$
This augmented constraint should be equivalent to (38) so that the second edge can be removed safely. A sufficient condition for this requirement is $T' > T_2$, which we will deduce from the dominance condition (34) and the condition $k_2 - k_1 > 0$ for (36). From (38) we have
$$T' \geq T_1 = (w_1 + w_3)/(k_1 + k_3) \Leftrightarrow k_1T' + k_3T' \geq w_1 + w_3. \quad (40)$$

If the edge pruning technique is applied in Algorithm 1 to calculate the minimum clock period, the new loop constraints are updated into $T_m$ using (11) gradually. At the moment of edge pruning, the variable $T_m$ used in the condition (36) is actually $T_c$ in (38), so that we can write the pruning condition in (36) as

$$T' \geq T_c > (w_2 - w_1)/(k_2 - k_1) \Leftrightarrow k_2 T' - k_1 T' > w_2 - w_1. \quad (41)$$

Adding both sides of (40) and (41) respectively, we have

$$k_2 T' + k_3 T' > w_2 + w_3 \Leftrightarrow T' > (w_2 + w_3)/(k_2 + k_3) = T_2. \quad (42)$$

Comparing (42) and (39) we know that (36) is a sufficient condition for the safe removal of the dominated edge.

According to (40)–(42) we can apply the pruning technique in Algorithm 1 to calculate $T_m$. When computing criticalities of gate delays from (17) to (30) using Algorithm 2, the minimum clock period $T_m$ has been calculated by Algorithm 1 in L5 of Algorithm 2. Therefore, we can not assume that $T_m$ used in (36) is equal to $T_c$. Instead, we use (25) to explain the pruning technique in criticality computation similar to (38)–(42). The condition $\max_{l \in L_e}\{T_l\} \geq T_m$ in (25) can be rewritten as

$$\max\{\max_{l \in L_e \setminus l_1, l_2}\{T_l\} - T_m, T_1 - T_m, T_2 - T_m\} \geq 0 \quad (43)$$

where $T_1$ and $T_2$ are the loop constraints from the loops formed by the two parallel edges and the third edge in Fig. 7, respectively. Let $d_c = \max\{\max_{l \in L_e \setminus l_1, l_2}\{T_l\} - T_m, T_1 - T_m\}$, and we have

$$d_c \geq T_1 - T_m = (w_1 + w_3)/(k_1 + k_3) - T_m \Leftrightarrow \quad (44)$$

$$(k_1 + k_3) d_c \geq (w_1 + w_3) - (k_1 + k_3) T_m. \quad (45)$$

From (36) we can deduce

$$(k_2 - k_1) T_m > w_2 - w_1. \quad (46)$$

Adding both sides of (45) and (46) we have

$$(k_1 + k_3) d_c > (w_2 + w_3) - (k_2 + k_3) T_m. \quad (47)$$

Because $T_m$ is the maximum of all loop constraints, it is no smaller than $\max_{l \in L_e \setminus l_1, l_2}\{T_l\}$ and $T_1$, so that we can deduce that $d_c$ is no greater than 0. Therefore, (47) can be written as

$$0 > (w_2 + w_3) - (k_2 + k_3) T_m \Leftrightarrow \quad (48)$$

$$T_m > (w_2 + w_3)/(k_2 + k_3) = T_2. \quad (49)$$

Thus we can state that with the condition (36) the loop containing the dominated edge with the weight $w_2 - k_2 T$ in Fig. 7 does not affect the minimum clock period and also does not contribute to the criticalities of gate delays.

The parallel pruning technique above is implemented in the function prune_edges() in Algorithm 1 and 2. During the iterations, the serial merge operations create new edges representing paths in the original graph. Therefore the new edge weights are balanced between consecutive stages and exhibit a tendency of being dominated by other parallel edges, so that they can be handled effectively by the parallel pruning technique. This can also be explained by the fact that in most cases a large combinational path delay in the circuit needs not to be compensated by flip-flop stages far away.

The pruning technique is applied in each iteration of Algorithm 2 to trim edges. The overall result is that only the critical part of the graph is unrolled during the iterations so that the runtime can be reduced. Moreover, the pruning technique also reduces redundant loops created by the serial merge operation, as exemplified in Fig. 8 where the edge weights are denoted as shown for simplification. In this graph, if node 1 is removed first by the serial merge operation, an edge with the weight $d_{21} + d_{13}$ is created from node 2 to node 3 in the graph on the right. If node 2 is removed thereafter, this edge is merged with the edge from 4 to 2 and a new edge with the weight $d_{41} + d_{12} + d_{21} + d_{13}$ will be created. The new edge is dominated by the edge from node 4 to node 3 with the weight $d_{41} + d_{13}$ because $d_{12} + d_{21}$ is no greater than 0 and this constraint has been captured by the self-loop at node 2. This example shows that although the serial merge operation does not miss any loop in the graph, it may create redundant edges. These edges are then pruned under the condition of parallel dominance discussed above, so that the number of the edges in the constraint graph does not increase unnecessarily.

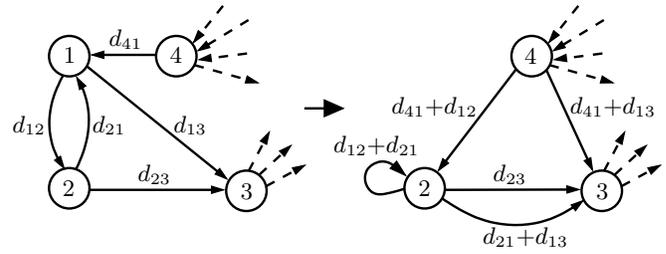

Fig. 8. Redundant edge created by serial merge operation. The edges between nodes 1 and 2 lead to a redundant edge from node 2 to node 3 after node 1 is removed by the serial merge operation.

*3) Merging tracing lists in computing criticalities*

In computing criticalities of gate delays using Algorithm 2, for each newly created edge we maintain an edge list to trace edges from which the new edge is created. Each time when a self-loop is removed, the loop constraint of each edge in the tracing list is updated in L21–L25 of Algorithm 2. The serial merge operation creates direct edges between predecessors and successors of the removed node. For such a new edge the traced edges in the lists of the two replaced edges are merged into the new tracing list, unless a self-loop is formed and the loop constraints for the traced edges are updated. The operation of merging edge tracing lists is implemented in the function update_tracing_lists() in L16–L18.

For parallel edges, we apply the pruning technique discussed above to reduce the number of edges in the function prune_edges(). Specifically, if the coefficients of $T$ in the edge weights of parallel edges are equal, we can compress them into one edge directly using the parallel merge operation parallel_merge() in L26–L30 of Algorithm 2. In this case, the two parallel edges with the weights $w_{ij_1} - k_1 T$ and $w_{ij_2} - k_2 T$ can be merged into a new edge. The new edge weight is calculated as $\max\{w_{ij_1}, w_{ij_2}\} - k_1 T$ because $k_1 = k_2$. When computing the criticalities, however, the parallel edges may have different edge tracing lists and the new edge weight can not be used for the traced edges from any of the lists, because $\max\{w_{ij_1}, w_{ij_2}\}$ is different from $w_{ij_1}$ and $w_{ij_2}$ which are calculated across different loops. To solve this problem, we maintain a random variable for each traced edge. In the case above, suppose we have an edge $e$ in the tracing list of the first merged edge. After the parallel operation, weight difference $\max\{w_{ij_1}, w_{ij_2}\} - w_{ij_1}$ is added to the tracing variable of $e$. This variable represents the difference between the weight of the new edge and the weight of the edge replaced by the parallel merge operation. When a self-loop is formed later, the loop constraint $T_l$ for edge $e$ can be recovered by the edge weight of the self-loop minus the accumulated weight difference for computing the criticalities of gate delays.

*4) Node order during graph transformation*

In Algorithm 1 and Algorithm 2 the next node for the serial merge operation is selected by the function select_node (). The order of the selected nodes may affect the performance significantly. If a node with $m$ predecessors and $n$ successors is removed by the serial merge operation, $m \times n$ new edges may be created. An extreme case is the root node, which has edges to and from all the other nodes. If the root node is removed using the serial merge operation, between any two nodes in the graph two new edges are constructed. In a graph with many nodes, it is impractical to process so many edges by further graph transformation.

In a complex graph, although an optimal node processing order that guarantees the minimal number of edges during graph transformation may exist, to find this optimal order is very difficult and, even if it is possible, consumes much runtime. In the proposed method, we select the next node for the serial merge operation heuristically. In each iteration we select the node with the smallest node connection, which is defined as $m_i \times n_i$ for node $i$, where $m_i$ is the number of predecessors of $i$ and $n_i$ the number of successors of $i$. With this heuristic method, the nodes that might lead to many new edges after their removal, for example, the root node, are processed later. The pruning techniques applied to the edges in the earlier iterations may reduce the number of the edges in the graph so that the overall runtime can be reduced. In implementation, an ordered list of all the nodes in the graph is maintained. Each time when a node is removed by the serial merge operation, the node connections of the predecessors and successors of the removed node are updated for refreshing the node order. Therefore, the next node for transformation is always at the head of the ordered list so that the runtime in node selection can be reduced.

*5) Discussions*

In Section III the timing constraints (2), (4) and (5) are represented in a constraint graph. The existence of a solution for the constraint set is equivalent to the condition that all loops in the graph have no positive accumulated weights. To capture the timing constraints from loops, we apply the graph transformation operations explained in Section III iteratively. The feasibility of this method can be explained using a sample in Monte Carlo simulation. For such a sample all the methods discussed above are still valid so that we can use them to calculate a minimum clock period for this sample. With all samples together, we have the distribution curve of the minimum clock period. From the statistical view, we can still apply the proposed method with the same node order and the same merge operations to calculate the minimum clock period. The only difference is that we should substitute statistical max and sum computations for the static counterparts. This is the same reasoning for statistical timing analysis of combinational circuits, where we use the same propagation algorithm but statistical computations.

The equivalence between the difference constraints and the graph representation is valid only when the ranges defined in (5) take continuous values. If the ranges of clock tuning elements are discrete as in many implementations, integer linear programming should be used to calculate the exact minimum clock period for each sample in Monte Carlo simulation. However, integer linear programming works differently with each sample of Monte Carlo simulation, for example, the different searching directions in finding the optimal value. Therefore, we cannot establish a closed-form formulation and thus cannot find a general method to perform fast statistical timing analysis in this case.

Instead of computing the exact minimum clock period $T_d$ directly, the result $T_m$ by assuming that all tuning elements have continuous ranges is a lower bound of the clock period of the case with discrete ranges, since all discrete configuration values are also feasible in the continuous configuration space. Assume that the interval of the discrete ranges of the clock tuning elements is $\theta$. Then $T_m + \theta$ is an upper bound for the minimum clock period in the discrete case, because for each solution of the continuous case we can move the delay of each tuning element to the nearest lower integer value to find a feasible discrete configuration that leads to the minimum clock period no larger than $T_m + \theta$. According to this discussion, we can conclude these two bounds as $T_m \leq T_d \leq T_m + \theta$. In reality, the interval of the ranges may be very small compared with the clock period, so that a good approximation accuracy can still be expected.

## V. EXPERIMENTAL RESULTS

The discussed algorithms were implemented in C++ and tested using a 2.67 GHz CPU. We used five large circuits, s5378 to s38584, from the ISCAS89 benchmarks and five other large circuits, mem_ctrl to des_perf, from TAU 2013 variation-aware timing analysis contest [40] for our experiments. Information about these circuits is shown in Table I, where $n_s$ is the number of flip-flops and $n_g$ the number of logic gates. The largest circuit in the experiments has more than 8K flip-flops and 86K logic gates. For experiments, we assumed that each flip-flop has a clock tuning element to test the efficiency of the proposed method with large constraint graphs. In practice, these tuning elements are selectively inserted considering circuit performance and area cost at the same time, e.g., using the methods in [16], [17]. The ranges of the clock tuning elements were set to 1/8 of the clock periods from the original circuits, roughly the range used in [14]. The logic gates in the circuits were mapped to a library from an industry partner. The standard deviations of transistor length, oxide thickness and threshold voltage were set to 15.7%, 5.3% and 4.4% of the nominal values, respectively [41]. The gate delays were generated using the method proposed in [3], in which spatial correlation from global and local variations is decomposed by principal component analysis (PCA). The resulting timing model is in a linear form of independent random variables. We used the method in [5] to compute the sum and maximum/minimum of random variables.

To verify the accuracy of the proposed method in computing

TABLE I
RESULTS OF STATISTICAL TIMING ANALYSIS AND CRITICALITY COMPUTATION

| Circuit | | SSTA accuracy | | | Runtime | | | Criticality>0.3 | | | Criticality>0.1 | | | Runtime | | |
|---|---|---|---|---|---|---|---|---|---|---|---|---|---|---|---|---|
| | $n_s$ | $n_g$ | $\mathcal{E}_\mu(\%)$ | $\mathcal{E}_\sigma(\%)$ | $\mathcal{E}_{T_{2\sigma}}(\%)$ | $t_p(s)$ | $t_m(h)$ | $r_t$ | $n_c$ | $n_m$ | $\mathcal{E}_c$ | $n_c$ | $n_m$ | $\mathcal{E}_c$ | $t_p(s)$ | $t_m(h)$ | $r_t$ |
| s5378 | 179 | 2779 | 0.55 | 1.18 | 0.05 | 0.10 | 0.39 | 14685 | 34 | 0 | - | 38 | 0 | - | 0.15 | 0.70 | 16851 |
| s9234 | 211 | 5597 | 0.13 | 0.42 | 0.10 | 0.17 | 0.99 | 21579 | 78 | 4 | 0.09 | 167 | 2 | 0.09 | 0.29 | 2.59 | 31733 |
| s13207 | 638 | 7951 | 0.12 | 0.23 | 0.04 | 0.16 | 1.75 | 39448 | 31 | 0 | - | 129 | 26 | 0.14 | 0.34 | 7.37 | 77800 |
| s15850 | 534 | 9772 | 0.62 | 1.04 | 0.16 | 0.43 | 2.94 | 24782 | 74 | 0 | - | 273 | 12 | 0.10 | 0.90 | 12.83 | 51407 |
| s38584 | 1426 | 19253 | 0.86 | 0.95 | 0.16 | 0.53 | 4.80 | 32832 | 74 | 0 | - | 110 | 1 | 0.04 | 1.29 | 47.18 | 132132 |
| mem_ctrl | 1065 | 10327 | 0.12 | 0.39 | 0.26 | 0.91 | 5.49 | 21700 | 37 | 2 | 0.06 | 44 | 0 | - | 3.20 | 59.02 | 66359 |
| usb_funct | 1746 | 14381 | 1.39 | 0.89 | 0.03 | 0.66 | 5.21 | 28575 | 56 | 4 | 0.05 | 94 | 0 | - | 1.90 | 67.67 | 128352 |
| ac97_ctrl | 2199 | 9208 | 1.19 | 0.35 | 0.21 | 0.60 | 5.45 | 32935 | 0 | 0 | - | 70 | 0 | - | 2.02 | 92.89 | 165359 |
| pci_bridge32 | 3321 | 12494 | 0.14 | 1.11 | 0.36 | 3.28 | 13.96 | 15296 | 33 | 0 | - | 74 | 4 | 0.03 | 8.79 | 350.75 | 143734 |
| des_perf | 8808 | 86020 | 0.75 | 0.13 | 0.15 | 4.81 | 55.48 | 41489 | - | - | - | - | - | - | 45.49 | - | - |

$n_s$: # of flip-flops; $n_g$: # of gates; $\mathcal{E}_\mu$: difference of mean in percentage; $\mathcal{E}_\sigma$: difference of standard deviation in percentage; $\mathcal{E}_{T_{2\sigma}}$: difference of yield at $2\sigma$ clock period; $t_p$: runtime of the proposed method in seconds; $t_m$: runtime of Monte Carlo simulation in hours; $r_t$: ratio of runtimes; $n_c$: # of gates with criticality above threshold; $n_m$: # of gates not captured by the proposed method; $\mathcal{E}_c$: maximum difference of the criticalities of the uncaptured gates.

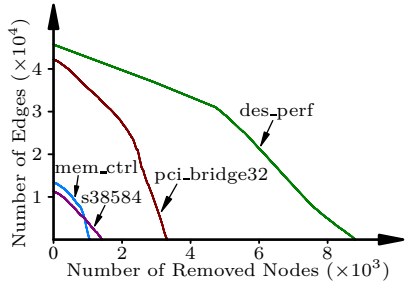

Fig. 9. Edge numbers during node removal. The numbers of edges exhibit monotonic decrease due to pruning and merging techniques.

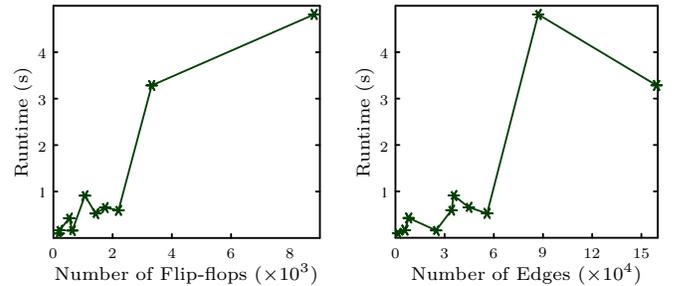

(a) Runtime vs. # of nodes  (b) Runtime vs. # of edges
Fig. 10. Runtime trend in relation to circuit size.

the minimum clock period, we ran Monte Carlo simulation with 10 000 samples. For each sample, the minimum clock period constrained by (2), (4) and (5) was computed using a linear programming solver [42]. The distribution formed by all the performance samples was compared with $T_m$ computed by the proposed method. The results are shown in Table I, where $\mathcal{E}_\mu$ is the relative error of the mean of the minimum clock period $T_m$, defined as $|\mu_{SSTA} - \mu_{MC}|/\mu_{MC}$, where $\mu_{SSTA}$ and $\mu_{MC}$ are the means of the minimum clock period computed by the proposed method and by Monte Carlo simulation, respectively. Similar to $\mathcal{E}_\mu$, $\mathcal{E}_\sigma$ shows the accuracy of the standard deviation of the clock period. $\mathcal{E}_{T_{2\sigma}}$ shows the relative yield error at the $2\sigma$ clock period from Monte Carlo simulation. From $\mathcal{E}_\mu$, $\mathcal{E}_\sigma$, and $\mathcal{E}_{T_{2\sigma}}$ we can see that the results of the proposed method have good accuracy and the predicted yields have no more than 0.5% error.

The major advantage of the proposed method is its efficiency. The runtimes of the proposed method working on different benchmark circuits are shown in Table I with $t_p$ in seconds, and the runtimes of Monte Carlo simulation are shown as $t_m$ in hours. The speedup ratios of the proposed method compared with Monte Carlo simulation are shown in the $r_t$ column. From this comparison, we can conclude that the proposed method is at least four orders of magnitude faster than Monte Carlo simulation. Since Monte Carlo simulation is the only existing method available for statistical timing analysis of circuits with clock tuning elements, this comparison demonstrates the advantage of the proposed method and its applicability to accelerate methods that depend on the results of statistical timing analysis, for example, in circuit optimization.

The proposed method unrolls the loops in the constraint graph using the serial merge operation, and parallel edges are pruned and merged to reduce the number of edges in the graph. In the worst case, the complexity of the algorithm is exponential in the numbers of nodes and edges in the constraint graph. The efficiency of the proposed method results from the techniques in Section IV, where edges are pruned during graph transformations. The effect of these heuristic techniques depends on the circuit structure and gate delays, so that the computational complexity cannot be presented in an accurate mathematical form. To demonstrate the efficiency of the proposed method, we show the trends of the edge numbers in the constraint graphs of several test cases in Figure 9. In all these cases, the numbers of edges actually decrease monotonically, since only the edges that affect the minimum clock period are kept in the graphs due to edge pruning, thus explaining the much shorter runtime compared with Monte Carlo simulation. To show the complexity trend of the proposed method, we illustrate the runtimes of processing circuits with different sizes regarding the numbers of nodes and edges in the constraint graphs in Fig. 10a and Fig. 10b, respectively. These diagrams show that the runtime of the proposed method increases with the circuit size, but still remains in the acceptable range. The complexity has a similar trend if different global and local variations are considered, because gate delays are usually represented in the same form, e.g., linear or quadratic polynomial of independent random variables.

Using clock tuning elements the performance of a circuit can be improved, as explained in Section II. However, the performance improvement is bounded because after the ranges of clock tuning elements reach a threshold the circuit performance is determined by the maximum average edge delay

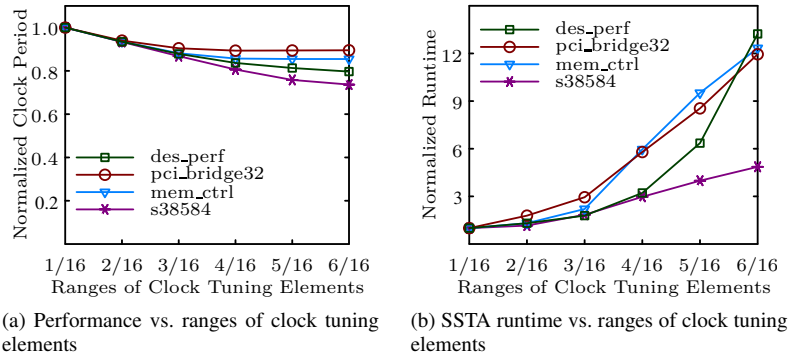

(a) Performance vs. ranges of clock tuning elements

(b) SSTA runtime vs. ranges of clock tuning elements

Fig. 11. Performance and runtime in relation to ranges of clock tuning elements. Circuit performance increases as the ranges are increased, but is bounded by the loops formed by paths across flip-flops exclusively. Runtime increases because the pruning techniques (33) and (37) become less effective with a decreased $T_m$.

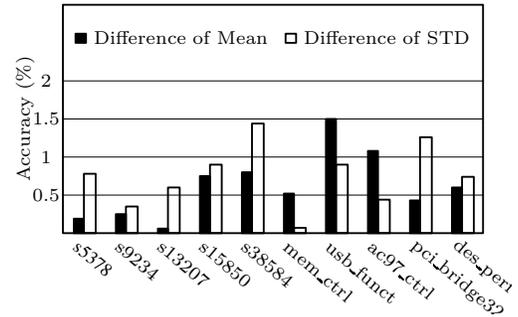

Fig. 12. Approximation accuracy of discrete Monte Carlo simulation. The results from SSTA for the circuits having clock tuning elements with continuous ranges are lower bounds for the discrete cases.

across loops exclusively. This maximum average edge delay does not change as the ranges of the clock tuning elements are enlarged, so that the circuit performance can not be improved by further time borrowing. To show the relation between circuit performance and ranges of clock tuning elements, we tested the circuit performances of some benchmark circuits by setting the ranges of tuning elements $r_i$ in (5) from $T_n/16$ to $6T_n/16$ with $T_n/16$ as interval, where $T_n$ is the clock period without considering the clock tuning elements. The trends of the mean values of the minimum clock periods are shown in Fig. 11a. From this diagram, we can see that the clock period of s38584 decreases as the ranges of clock tuning elements are enlarged. But the clock period of pci_bridge32 nearly has no change when the ranges of clock tuning elements are larger than $3T_n/16$, because in this case the constraints from loops across flip-flops dominate the circuit performance. In Fig. 11b the trends of the runtimes of the proposed method with respect to different ranges of clock tuning elements are also shown. It is obvious that the runtimes increase as the ranges of clock tuning elements are enlarged, because the pruning techniques (33) and (37) become less effective with a decreased $T_m$. However, even with this increase, the absolute runtime of the proposed method is still small. For example, the analysis of the largest case des_perf finished within 50 seconds but the corresponding Monte Carlo simulation did not produce the result even in days. Therefore, the proposed method can be used to evaluate the relation between the minimum clock period and the ranges of clock tuning elements efficiently, so that designers have the chance to evaluate tradeoffs between performance and die size of the tuning elements.

To verify the proposed criticality computation, we sampled the constraint graph in each iteration of Monte Carlo simulation. Then we calculated the distances between nodes in the constraint graph using the Bellman-Ford algorithm. After this, each edge was checked whether the loop across it determines the minimum clock period computed by linear programming. The criticalities from Monte Carlo simulation and the proposed method are compared and the results are shown in Table I. Owing to the approximation in the statistical computations of SSTA engines, the criticalities can not be calculated accurately. As pointed out in [25], the skewness of the distribution, which is not considered by many SSTA engines, may cause large inaccuracy in criticality computation.

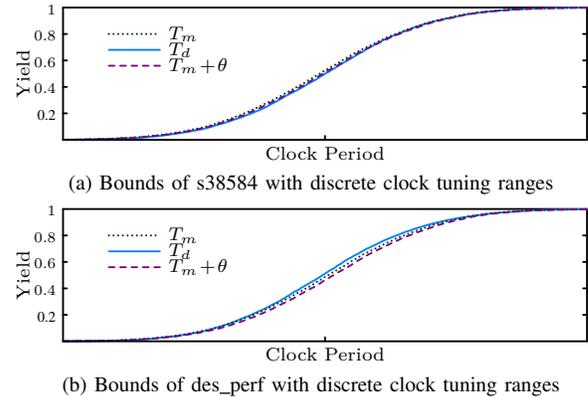

(a) Bounds of s38584 with discrete clock tuning ranges

(b) Bounds of des_perf with discrete clock tuning ranges

Fig. 13. Approximation and bounds of minimum clock periods of circuits with discrete clock tuning ranges. The upper and lower bounds have good accuracy of approximation, but are not bounds exactly due to the approximation in statistical computations.

Especially when the delays of critical paths are compared with the minimum clock period of the circuit, the criticality is very sensitive to the inaccuracy of the statistical approximations. Nevertheless, because the purpose to compute criticalities is to select the gates for optimization, we compare the sets of critical gates selected by Monte Carlo simulation and the proposed method. In Table I the columns $>0.3$ and $>0.1$ show the numbers of gate delays with criticalities larger than $0.3$ and $0.1$, respectively. The comparison of criticalities for des_perf was not fulfilled in the experiment due to the unaffordable runtime in Monte Carlo simulation. In Table I $n_c$ is the number of gates identified by Monte Carlo simulation, and $n_m$ is the number of gates which are not captured by the proposed method. $\mathcal{E}_c$ is the maximum difference between the criticalities of gates which are not captured by the proposed method. For example, for s9234, in the 78 gates with criticalities larger than $0.3$ four gates are not captured. In these four missing gates, the maximum criticality difference compared with the criticalities computed by Monte Carlo simulation is $0.09$. From this comparison, we can conclude that the proposed method can capture most of the critical gates, but it may still miss some due to the approximation in statistical computations, though the difference between the criticalities is not large. The runtime comparison for computing criticalities is shown in the last three columns of Table I. The acceleration ratio of the proposed method to Monte Carlo simulation is still remarkable, particularly for the large benchmark circuits.

In Section IV we have discussed the upper and lower bounds of the minimum clock period $T_d$ for circuits containing clock tuning elements with discrete ranges. If we assume the ranges are continuous and apply the proposed method, the resulting minimum clock period $T_m$ is a lower bound of $T_d$. In addition, if we increase $T_m$ by $\theta$ which is the discrete interval of the ranges of the clock tuning elements, we then create an upper bound for $T_d$. In Fig. 12 we show the differences of means and standard deviations of $T_m$ and $T_d$. Here the discrete adjustable range has eight steps. These differences shown in the y-axis in percentage demonstrate a reasonable approximation of $T_m$ to $T_d$ generally. In Fig. 13a and 13b the cumulative distribution functions of $T_m$ as the lower bound, $T_d$ as the result of Monte Carlo simulation, and $T_m + \theta$ as the upper bound for circuits s38584 and des_perf are shown, respectively. In these two comparisons, both bounds are lower bounds for s38584 and upper bounds for des_perf, due to the approximation in the max and sum computations in statistical timing analysis. But these bounds all exhibit a reasonable approximation accuracy.

## VI. Conclusion

In this paper, we propose a fast method to compute the minimum clock periods for circuits with post-silicon clock tuning elements. The delays of these elements can be adjusted for each individual chip after manufacturing to achieve the maximum performance. The proposed method applies serial merge operations to unroll the loops in the constraint graph so that non-positive loop constraints can be captured by self-loops. Parallel merge operations and pruning techniques are applied to trim edges during iterations to reduce runtime. Criticalities of logic gates are also calculated by tracing the edges on critical loops. Experimental results confirm that the propose method is faster than Monte Carlo simulation by several orders of magnitude while still maintaining good accuracy.

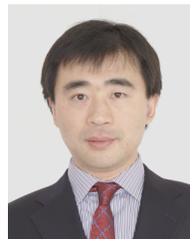

**Bing Li** received the bachelor's and master's degrees in communication and information engineering from Beijing University of Posts and Telecommunications, Beijing, China, in 2000 and 2003, respectively, and the Dr.-Ing. degree in electrical engineering from Technische Universität München (TUM), Munich, Germany, in 2010. He is currently a researcher with the Institute for Electronic Design Automation, TUM. His research interests include timing and power analysis and emerging systems.

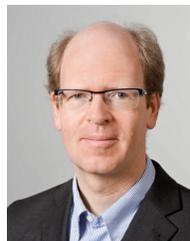

**Ulf Schlichtmann** (S'88–M'90) received the Dipl.-Ing. and Dr.-Ing. degrees in electrical engineering and information technology from Technische Universität München (TUM), Munich, Germany, in 1990 and 1995, respectively. He was with Siemens AG and Infineon Technologies AG, Munich, from 1994 to 2003, where he held various technical and management positions in design automation, design libraries, IP reuse, and product development. He has been with TUM as a Professor and the Head of the Institute for Electronic Design Automation, since 2003. He served as the Dean of the Department of Electrical Engineering and Information Technology, TUM, from 2008 to 2011. His current research interests include computer-aided design of electronic circuits and systems, with an emphasis on designing reliable and robust systems.